# Deep Learning Classification of Polygenic Obesity using Genome Wide Association Study SNPs

Casimiro A. Curbelo Montañez
*Department of Computer Science*
*Liverpool John Moores University*
Liverpool, UK
c.a.curbelomontanez@ljmu.ac.uk

Paul Fergus
*Department of Computer Science*
*Liverpool John Moores University*
Liverpool, UK
p.fergus@ljmu.ac.uk

Almudena Curbelo Montañez
*Hospital Nuestra Señora de Guadalupe*
La Gomera, Spain
contact@acurbelo.com

Carl Chalmers
*Department of Computer Science*
*Liverpool John Moores University*
Liverpool, UK
c.chalmers@ljmu.ac.uk

*Abstract*— **In this paper, association results from genome-wide association studies (GWAS) are combined with a deep learning framework to test the predictive capacity of statistically significant single nucleotide polymorphism (SNPs) associated with obesity phenotype. Our approach demonstrates the potential of deep learning as a powerful framework for GWAS analysis that can capture information about SNPs and the important interactions between them. Basic statistical methods and techniques for the analysis of genetic SNP data from population-based genome-wide studies have been considered. Statistical association testing between individual SNPs and obesity was conducted under an additive model using logistic regression. Four subsets of loci after quality-control (QC) and association analysis were selected: P-values lower than $1\times10^{-5}$ (5 SNPs), $1\times10^{-4}$ (32 SNPs), $1\times10^{-3}$ (248 SNPs) and $1\times10^{-2}$ (2465 SNPs). A deep learning classifier is initialised using these sets of SNPs and fine-tuned to classify obese and non-obese observations. Using a deep learning classifier model and genetic variants with P-value $< 1\times10^{-2}$ (2465 SNPs) it was possible to obtain results (SE=0.9604, SP=0.9712, Gini=0.9817, LogLoss=0.1150, AUC=0.9908 and MSE=0.0300). As the P-value increased, an evident deterioration in performance was observed. Results demonstrate that single SNP analysis fails to capture the cumulative effect of less significant variants and their overall contribution to the outcome in disease prediction, which is captured using a deep learning framework.**

## I. Introduction

Obesity prevalence has increased for several decades, and so has its impact on morbidity and mortality. According to the World Health Organisation (WHO), overweight and obesity comprise the fifth leading risk factor for mortality, resulting in approximately 2.8 million deaths globally every year [1]. Individuals suffering from obesity are at higher risk of diabetes, cardiovascular disease, cancer and many other health problems [2], significantly limiting their life expectancy. The effect of obesity and its consequences affect the annual direct cost of the National Health Service (NHS) which was estimated to be approximately £5.1 billion in 2006-07 [3]. Childhood obesity is also a major public health problem [4], with obesity prevalence increasing from 11.1% and 12.2% to 16.8% and 15.2% in boys and girls respectively in the UK from 1995 to 2009 [5]. Obese children are more likely to become obese adults, with associated health problems and consequential costs [6].

It is very likely that the genetic makeup of an individual may influence his/her susceptibility to complex disease traits, such as obesity. Observations of variability in obesity susceptibility among individuals, or communities, exposed to the same environmental risk factors suggests that genetic differences have an appreciable role in the observed individuals' variation in body weight and obesity susceptibility. This is further supported by twin and family studies which found that variation in body mass index (BMI) was largely due to heritable genetic differences, with heritability estimates in adults ranging from 50% to 70% [7], [8].

Genome-wide association studies (GWAS) reveal variants at genomic loci that are associated with complex traits in the population, providing the opportunity to explore the genetic basis of complex disorders. In these studies, a large number of genetic variants or single nucleotide polymorphisms (SNPs) are tested for association with the trait of interest or common diseases, such as obesity, diabetes or coronary artery disease among others [9], [10]. SNPs are the most common type of genetic variation, resulting from a single base substitution in the deoxyribonucleic acid (DNA) sequence [11]. Results from GWAS reveal SNPs that serve as candidate biomarkers for genes that might indicate the existence of complex diseases in individuals. This approach is based on single-locus analysis where each SNP is independently tested for association with a phenotype of interest, omitting the existence of interactions between loci.

Significant associations between the investigated trait and a SNP are termed genome-wide significant. These associations can be measured using Bonferroni correction, a highly conservative method designed to minimize type 1 errors in multiple testing studies. Currently, genome-wide significant threshold of $5\times10^{-8}$ has been accepted as strong evidence for association [12].

Filtering approaches to reduce and eliminate redundant information in the set of genotyped SNPs have been used [13]. This modifies the data representation space, facilitating the detection of non-linear interactions among the variables. Performing SNP selection based on arbitrary significance threshold by calculating a test statistic for each marker separately and evaluating all possible interactions in the filtered subset is one way of reducing computational complexity [14]. Nonetheless, there are other ways to reduce the number of variants for subsequent analysis that do not require a locus to have significant marginal effects [15]. Machine learning techniques such as Neural Networks (NNs), Random Forest (RF) and Cellular Automata (CAs) have been integrated with constructive induction algorithms such as Multifactor Dimensionality Reduction (MDR) to detect epistasis interactions [16]. Although methods such as MDR exhaustively consider every possible combination of SNPs up to a defined order, they still suffer from computational complexity issues in large candidate gene studies and genome-wide association studies. Thus, successful data analysis is currently dependent on the combination of traditional and novel data analysis methods.

Reaching a better understanding behind the causes of obesity and related comorbidities, including cardiovascular disease and type 2 diabetes mellitus (T2DM) is one of the main goals of genetic investigation of obesity. Furthermore, the interpretation and translation of such knowledge opens new opportunities to introduce personalised medicine to obese patients, enabling more specific diagnosis of the causal factors underlying obesity. The importance of GWAS is advancing scientific understanding of disease mechanisms as well as providing starting points and potential opportunities to improve the development of medical treatments or prevention therapies [17].

In this paper, we combined association results from GWAS with a deep learning (DL) framework to test the predictive capacity of statistically significant SNPs associated with obesity phenotype. Our approach demonstrates the potential of deep learning as a powerful framework for extending GWAS analysis that can capture information about SNPs and the important interactions between them. Basic statistical methods and techniques for the analysis of genetic SNP data from population-based genome-wide studies have been considered, particularly logistic regression. Four subsets of loci after quality-control (QC) and association analysis were selected: P-values lower than $1 \times 10^{-5}$ (5 SNPs), $1 \times 10^{-4}$ (32 SNPs), $1 \times 10^{-3}$ (248 SNPs) and $1 \times 10^{-2}$ (2465 SNPs). A deep learning classifier is initialised using these sets of SNPs and fine-tuned to classify obese and non-obese observations.

The remainder of this paper is organised as follows. Section 2 describes the Materials and Methods used in the study. The results are presented in Section 3 and discussed in Section 4 before the paper is concluded and future work is presented in Section 5.

## II. MATERIALS AND METHODS

In this section, the data used in the study is introduced. In addition, QC, association analysis and deep learning are discussed.

### A. Data Collection

The case and control data utilized in this study were requested from the database of Genotypes and Phenotypes (dbGaP) [18]. Participants considered are part of the MyCode Community Health Initiative (MyCode) project, a powerful platform for translational research [19]. MyCode was created as a central biorepository to collect blood and DNA samples from a representative cohort of patients from the Geisinger Health System (GHS), an integrated health care delivery system which provides services to participants resident in Pennsylvania. Samples and molecular data generated by MyCode have been used in numerous research studies, including the electronic Medical Records and Genomics (eMERGE) Network. This network represents a collaboration of institutions with biobanks linked to electronic medical records (EMRs), supported by the National Human Genome Research Institute (NHGRI) [20].

Cases and controls provided by dbGaP have been extracted as part of different study cohorts from the Geisinger MyCode project. Control patients from the eMERGE Geisinger eGenomic Medicine (GeM) - My-Code Project Controls (dbGaP study accession phs000381.v1.p1), were eligible if they were primary patients of a Geisinger Clinic with non-urgent visits to the clinic. A subset of 1,231 unique samples were genotyped using Illumina HumanOmniExpress-12 v1.0 arrays, and used as population controls for the eMERGE Genome-Wide Association Studies of Obesity project (dbGaP study accession phs000408.v1.p1). Case samples were part of a cohort of primary Caucasian patients from the Geisinger Clinic with extreme obesity who have undergone bariatric surgery. This time, a subset of 962 unique samples with a mean BMI of 49.17 ($\pm$ 8.83 SD) were genotyped using Illumina HumanOmniExpress-12 v1.0 arrays. All study participants provided written consent prior to study enrolment as part of MyCode DNA biobank.

The control group includes 448 females and 743 males with a mean age of 66.74 ($\pm$ 13.95 SD), while cases are composed of 788 females and 174 males with a mean age, at surgery, of 46.42 ($\pm$ 11.26 SD). Therefore, the whole case-control set contains a total of 2,193 participants of which 917 are males and 1,236 are females. Each participant contains 594,034 markers. Furthermore, 99.5% of the participants belong to a white ethnical background (Caucasians) as shown in Table I.

### B. Quality Control

To conduct association analyses, only those individuals reported to be Caucasian (white) were selected to reduce potential bias due to population stratification [21]. In addition, a series of analyses were also conducted to identify potentially problematic samples and SNPs. This step, known as data quality-control has become an imperative step prior to any GWAS analysis. Thus, QC of the genotyped data and filtering procedures were performed on individuals and then on markers, following standard QC protocols and guidance from [22].

TABLE I. CASE-CONTROL ETHNICITY

| Ethnicity | Case | Control | Total samples |
|---|---|---|---|
| White | 960 | 1,223 | 2,183 |
| AM* | 1 | 5 | 6 |
| AI/AN** | 0 | 1 | 1 |
| Hispanic*** | 0 | 1 | 1 |
| NH**** | 0 | 1 | 1 |
| Unknown | 1 | 0 | 1 |

*Black or African American*

**American Indian/Alaska Native*

***Hispanic or Latino*

****Native Hawaiian or other Pacific Islands*

The first step was to filter out and remove data samples with discordant sex. Related or duplicated samples were removed using Identity by Descent (IBD) coefficient estimates (IBD > 0.185). Principal component analysis (PCA) was performed to identify outliers and hidden population structure using EIGENSTRAT [23]. Genetic markers (SNPs) were excluded from analysis if the minor allele frequency (MAF) was lower than 5%, the Hardy-Weinberg Equilibrium P-value was lower than $1\times10^{-4}$ in control subjects, or if the genotype missing rate was higher than 0.001%. To remove data redundancy due to Linkage Disequilibrium (LD), each individual chromosome was scanned using a moving window with window size set to 50 SNPs with a step length of 5 SNPs. Furthermore, LD cut-off was set to 0.2. After QC, 1,997 individuals (879 cases and 1,118 controls) and 240,950 genetic variants remained for subsequent analysis.

QC analysis and successive association tests were conducted using PLINK [24] and the language and environment for statistical computing and graphics, R [25]. A Linux Ubuntu version 16.04 LTS based machine, with 64GiB of Memory and an Intel® Core™ i7-7700K CPU @ 4.20GHz x 8, was utilized to conduct QC filtering and deep learning analysis with H2O [26].

*C. Association Analysis*

Association analysis in case-control studies compares the frequency of alleles or genotypes at genetic loci (SNP) between cases and controls from a given population [27]. It starts by testing each SNP sequentially with the null hypothesis of no association. Thus, in association analysis, a series of single locus statistical tests explore each SNP separately and their likely association with a phenotype.

In the current study, statistical association testing between individual SNPs and obesity was conducted under an additive model using logistic regression, which is a widespread method for single SNP examination. A logistic function was used to predict the probability of a case given a genotype class. Furthermore, logistic regression allows for adjustment for clinical covariates and can provide adjusted odds ratios. Other genetic models are available, although logistic regression is the preferred approach.

In an additive genetic model, each copy of the minor allele increases the risk by the same amount. For example, if the risk is r for Aa, then there is a risk of 2r for AA. The advantage of examining only additive models is its capacity to detect both additive and dominant effects with reasonable power.

Let $G_{ij}$ be the genotypes AA, Aa, and aa where $j$ is the number of SNPs (j = 1, 2, ..., m), $i$ is the individuals (i = 1, 2, ..., n) and a and A are the minor and major allele respectively. Let $Y \in \{0,1\}$ be a binary phenotype for case/control status and $G_{ij} \in \{0,1,2\}$ be a genotype at the typed locus, where 0, 1 and 2 represent homozygous major allele *AA*, heterozygous allele *Aa*, and homozygous minor allele *aa* respectively. Logistic regression modelling is therefore defined as [28]:

$$logit(p_{ij}) = log\left(\frac{p_{ij}}{1-p_{ij}}\right) = \beta_0 + \beta_1 G_{ij}, \quad (1)$$

where $p_{ij}$ is the disease risk of the $j^{th}$ SNP on the $i^{th}$ individual.

*D. Deep Learning*

In this study, a multi-layer feedforward neural network is implemented based on the formal definitions in [29]. Labelled training samples $(x^{(i)}, y^{(i)})$ from case-control genetic data are considered for a supervised learning problem. A complex non-linear hypotheses $h_{W,b}(x)$ is defined using the neural network, with parameters $W,b$ fitted to our data.

Taking $(x_1, x_2,...x_n)$ and a +1 intercept term as input, computational units or neurons output

$$h_{W,b}(x) = f(W^T x) = f(\sum_{i=1}^{n} W_i x_i + b) \quad (2)$$

where $f: \mathbb{R} \mapsto \mathbb{R}$ represents the activation function.

Input, hidden and output layers make up the network structure where $n_l$ represent the number of layers, $L_1$ the input layer and $L_{n_l}$ the output layer. Several parameters constitute the neural network,

$$(W,b) = (W^{(1)}, b^{(1)}, W^{(n)}, b^{(n)}) \quad (3)$$

where $W_{ij}^{(l)}$ is the weight of the connection between unit $j$ in layer $l$, and unit $i$ in layer $l+1$. Parameter $b_i^{(l)}$ (intercept node), associated with unit $i$ in layer $l+1$, is introduced to counteracts the problem associated with input patterns that are zero. Additionally, the number of nodes in layer $l$ is denoted by $s_l$ without taking $b_i^{(l)}$ into consideration. Thus, $a_i^{(l)}$ refers to the activation (output) of node $i$ in layer $l$ of the network. The neural network defines $h_{W,b}(x)$ which outputs a real number based on a given set of parameters $W,b$.

Given a training set $\{(x^{(1)}, y^{(1)}),..., (x^{(m)}, y^{(m)})\}$ of $m$ samples, the neural network is trained using gradient descent and the overall cost function is defined as

$$J(W,b) = \left[\frac{1}{m}\sum_{i=1}^{m} J(W,b,x^{(i)}, y^{(i)})\right] + \frac{\lambda}{2}\sum_{l=1}^{n_l-1}\sum_{i=1}^{s_l}\sum_{j=1}^{s_l+1}(W_{ji}^{(l)})^2$$
$$= \left[\frac{1}{m}\sum_{i=1}^{m}(\frac{1}{2}//h_{W,b}(x^{(i)}) - y^{(i)}//^2)\right] + \frac{\lambda}{2}\sum_{l=1}^{n_l-1}\sum_{i=1}^{s_l}\sum_{j=1}^{s_l+1}(W_{ji}^{(l)})^2 \quad (4)$$

where the first term is the average sum of squared errors and the second, a weight decay or regularization term that helps prevent overfitting by reducing the magnitude of the weights. Hence, relative importance of the two expressions is controlled with the weight decay parameter $\lambda$.

Each parameter $W_{ij}^{(l)}$ and $b_i^{(l)}$ are initialized to a random value close to zero before the training, as it prevents hidden layer units learning the same function of the input. Following random initialization, gradient descent updates the $W,b$ as follows:

$$W_{ij}^{(l)} := W_{ij}^{(l)} - \alpha \frac{\partial}{\partial W_{ij}^{(l)}} J(W,b)$$
$$b_i^{(l)} := b_i^{(l)} - \alpha \frac{\partial}{\partial b_i^{(l)}} J(W,b) \quad , \quad (5)$$

where α is the learning rate.

The backpropagation algorithm (see Algorithm 1) efficiently computes the partial derivatives $\frac{\partial}{\partial W_{ij}^{(l)}} J(W,b;x,y)$ and $\frac{\partial}{\partial b_i^{(l)}} J(W,b;x,y)$ of the cost function $J(W,b;x,y)$ for a single sample.

An error term $\delta_i^{(l)}$ is computed for each node $i$ in layer $l$ to quantity the node's contribution to errors that occurred in the output. The error term $\delta_i^{(n_l)}$ for an output node ($n_l$ is the output layer), measures the difference between the activation and the true target value for an output node of the network. Conversely, hidden units compute $\delta_i^{(l)}$ by mean of a weighted average of the error terms of the nodes that use $a_i^{(l)}$ as input.

The derivatives for the overall cost function can be calculated once the partial derivatives have been computed. Hence:

$$\frac{\partial}{\partial W_{ij}^{(l)}} J(W,b) = \left[ \frac{1}{m} \sum_{i=1}^{m} \frac{\partial}{\partial W_{ij}^{(l)}} J(W,b;x^{(i)},y^{(i)}) \right] + \lambda W_{ij}^{(l)},$$
$$\frac{\partial}{\partial b_i^{(l)}} J(W,b) = \frac{1}{m} \sum_{i=1}^{m} \frac{\partial}{\partial b_i^{(l)}} J(W,b;x^{(i)},y^{(i)}) \quad (6)$$

where weight decay affects $W$ but not $b$.

Next, the gradient descent algorithm can be described. See Algorithm 2, where $\Delta W^{(l)}$ is a matrix with dimension equal to $W^{(l)}$, and $\Delta b^{(l)}$ is a vector with dimension equal to $b^{(l)}$.

Gradient descent is used to reduce our cost function $J(W,b)$ before training the neural network used in this study for classification purposes.

*E. Performance Metrics*

In this study, sensitivity (SE) and specificity (SP) are used to quantify how effectively a classifier identifies control and case instances. Sensitivity describes the true positive rate (Controls – Non-obese) and Specificity the true negative rate (Cases – obese).

The area under the receiver operating characteristic (ROC) curve (AUC) is used as a performance measure to evaluate the discrimination ability of the deep learning classifier. AUC measures the probability that test values from a randomly selected pair of case and control samples are correctly ranked and is thus a convenient global measure for the quantification of classification accurateness.

Logarithmic Loss (logloss) is a classification loss function which provides a measure of accuracy for a classifier by penalising false classifications. Minimising the logloss is correlated with accuracy, as one increases the other decreases. The logloss for a binary class classifier is defined by:

$$logloss = -\frac{1}{N} \sum_{i=1}^{N} \left[ y_i log(p_i) + (1 - y_i) log(1 - p_i) \right] \quad (7)$$

where $N$ is the number of samples, $p_i$ is the probability of $i^{th}$ sample belonging to class $C_1$ and $y_i$ is the actual label of the $i^{th}$ sample, which could be either 0 or 1. In case of misclassification, logloss values are progressively larger, whereas logloss for models that classify all instances correctly will be 0. Thus, the robustness of the model increases by minimising this value.

---

**Algorithm 1.** Backpropagation Algorithm

1: Perform a forward pass and compute activations for $L_2, \ldots, Ln_l$
2: **for** output unit $i$ in layer $n_l$, **do**
3: $\delta_i^{(n_l)} = \frac{\partial}{\partial z_i^{(n_l)}} \frac{1}{2} ||y - h_{W,b}(x)||^2 = -(y_i - a_i^{(n_l)}) \cdot f'(z_i^{(n_l)})$
4: **end for**
5: **for** $l = n_l - 1, \ldots, 2$, **do**
6: **for** node $i$ in layer $l$, **do**
7: $\delta_i^{(l)} = \left( \sum_{j=1}^{S_l+1} W_{ji}^{(l)} \delta_j^{(l+1)} \right) f'(z_i^{(l)})$
8: **end for**
9: **end for**
10: Compute the desired partial derivatives:
11: $\frac{\partial}{\partial W_{ij}^{(l)}} J(W,b;x,y) = a_j^{(l)} \delta_i^{(l+1)}$
12: $\frac{\partial}{\partial b_i^{(l)}} J(W,b;x,y) = \delta_i^{(l+1)}$

---

**Algorithm 2.** Gradient Descent

1: Set $\Delta W^{(l)} := 0$, $\Delta b^{(l)} := 0$ (matrix/vector of zeros) for all $l$.
2: **for** $i = 1, \ldots, m$, **do**
3: Use backpropagation to compute $\nabla_{W^{(l)}} J(W,b;x,y)$ and $\nabla_{b^{(l)}} J(W,b;x,y)$.
4: Set $\Delta W^{(l)} := \Delta W^{(l)} + \nabla_{W^{(l)}} J(W,b;x,y)$
5: Set $\Delta b^{(l)} := \Delta b^{(l)} + \nabla_{b^{(l)}} J(W,b;x,y)$
6: **end for**
7: Update the parameters:
8: $W^{(l)} := W^{(l)} - \alpha \left[ (\frac{1}{m} \Delta W^{(l)}) + \lambda W^{(l)} \right]$
9: $b^{(l)} := b^{(l)} - \alpha \left[ \frac{1}{m} \Delta b^{(l)} \right]$

A common performance metric utilised to measure the average sum of the square difference between actual values and predicted values is the Mean Squared Error (MSE). The standard definition of MSE is defined as:

$$MSE = \frac{1}{n}\sum_{i=1}^{n}(y^{(i)} - \hat{y}^{(i)})^2 \quad (8)$$

where $(y^{(i)} - \hat{y}^{(i)})$ is termed residual, and the aim of MSE is to minimise the residual sum of squares. MSE values close to 0 indicates that the model correctly classifies all class instances. Progressively larger values of MSE indicates misclassification.

The Gini coefficient is occasionally used in binary classification studies and is defined as being the area between the diagonal line and the ROC curve:

$$Gini = 2*AUC - 1 \quad (9)$$

The Gini coefficient quantifies dispersion among values of a frequency distribution. Gini values close to 1 indicates a good model, whereas a coefficient of 0 indicates that the features (SNPs) have no predictive capacity.

## III. RESULTS

Genotype data for 960 cases and 1,223 control subjects was analysed. After QC, 1,997 individuals and 240,950 SNP genotypes remained. These were then used to conduct association analysis of obesity traits.

Association tests between SNP genotypes and obesity traits were performed using the PLINK software version 1.09. Logistic regression was adjusted using Genomic Control (GC) to control population structure. Correction for multiple testing was also considered using the Bonferroni correction [12]. Bonferroni correction adjusts a commonly used alfa value from $\alpha = 0.05$ to $\alpha = (0.05/n)$ where $n$ is the number of statistical tests conducted. Hence, we performed analysis to assess associations of all Illumina HumanOmniExpress-12 v1.0 arrays SNPs with obesity phenotype, in which we adjusted for the total number of SNPs tested, defining the significance cut-off as P-value < $2.1 \times 10^{-7}$ after Bonferroni's correction (P-value = 0.05/240,950).

Fig. 1 shows the P-values for this study by means of a Manhattan plot. P-values in $-\log_{10}$ scale are distributed in the $y$ axis while the physical position of the SNPs along chromosomes is represented in the $x$ axis. The smallest P-values suggest potential disease-related SNPs. Bonferroni corrected significant threshold and suggestive threshold of association are represented in Fig. 1 using red and blue lines, respectively.

None of the SNPs identified in logistic analysis reached the Bonferroni level of significance (P-value < $2 \times 10^{-7}$ - red horizontal line in Fig. 1); however, five SNPs were suggestive of association (P < $1 \times 10^{-5}$ - blue horizontal line in Fig. 1). All suggestive association signals with P-value < $1 \times 10^{-5}$ are shown in Table II.

Since individual SNPs tend to have small effect sizes, the effect of the SNPs with higher significance (lower P-values) is investigated. Thus, modified suggestive thresholds has been considered to increase the number of SNPs for investigation in this study, based on our previous work [30], [31]. These SNPs capture the linear interactions between SNPs and phenotype but not the existing cumulative interactions between them.

### A. Baseline Deep Learning Network

After QC and association analysis using logistic regression, four different sets of SNPs were derived by the application of different P-value thresholds. The suggestive threshold of association ($1 \times 10^{-5}$) was considered, in addition to $1 \times 10^{-4}$, $1 \times 10^{-3}$, and $1 \times 10^{-2}$. Therefore, four sets of SNPs (5, 32, 248 and 2465 SNPs) are used to fit a deep learning network, based on a multi-layer feedforward artificial neural network (ANN) that is trained with stochastic gradient descent using back-propagation. MSE, Logloss, AUC, Gini, Sensitivity and Specificity values are used to measure the performance of each model. The data set is split randomly into training (60%), validation (20%) and testing (20%).

#### 1) Hyperparameters selection.

The four models considered were trained with global and specific hyperparameters. To maximize the predictive performance of our models, we focused on tuning the network architecture and the regularization parameters, while we used adaptive learning rate.

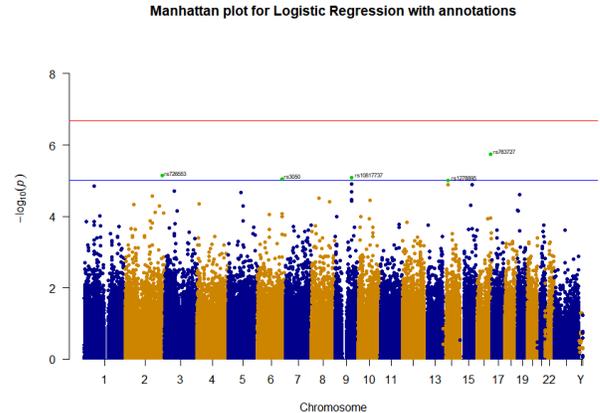

Fig. 1. Manhattan plot for logistic test adjusted GC.

Early stopping is adopted to avoid overfitting. The model stops when logloss does not improve by at least 1% (stopping tolerance) for two scoring events (stopping rounds). A maximum on the sum of the squared incoming weights into any one neuron was set to 10, this can help improve stability for Maxout or Rectifier. The learning rate is configured to 0.005 with rate annealing set to $1 \times 10^{-6}$ and rate decay set to 1. The learning rate is a function of the difference between the predicted value and the target value. This is a delta value available at the output layer. The output at each hidden layer is corrected using backpropagation. Rate annealing is utilised during this process to reduce the learning rate to freeze into local minima. Rate decay simply controls the change of learning rate across layers.

Additionally, more specific tuning parameters were applied to each model for training purposes. In the first case (5 SNPs), a deep learning classification model with 2 hidden layers with 20 neurons each was used. The activation function used throughout the network is MaxoutWithDropout. To add stability and improve generalization, L1 and L2 regularisation values were set to $7.1 \times 10^{-5}$ and $9.6 \times 10^{-5}$ respectively. In the second (32 SNPs) and third (248 SNPs) cases, 2 hidden layers with 50 neurons each, a TanhWithDropout activation function, and L1=$3.0 \times 10^{-6}$ and L2=$6.5 \times 10^{-5}$ values were used to train the classifier. Finally, to train the final model (2465 SNPs), 2 hidden layers with 10 neurons each, a RectifierWithDropout activation function, and L1=$9.6 \times 10^{-6}$ and L2=$2.8 \times 10^{-5}$ values were used.

TABLE II. SUGGESTIVE RESULTS (P < 1 x 10⁻⁵) FOR GWAS ON OBESITY

| SNP | Chromosome number | Position | Closest Gene | SNP-gene distance | P-value | Odds ratio |
|---|---|---|---|---|---|---|
| rs763727 | 16 | 83342301 | CDH13 | 0 | $1.821 \times 10^{-06}$ | 1.3770 |
| rs726553 | 2 | 226016494 | Intergenic | - | $7.330 \times 10^{-06}$ | 1.3530 |
| rs10817737 | 9 | 100306267 | TMOD1 | 0 | $8.319 \times 10^{-06}$ | 1.3490 |
| rs3050 | 6 | 150923115 | PLEKHG1 | 0 | $9.061 \times 10^{-06}$ | 0.6803 |
| rs1278895 | 14 | 32400170 | Intergenic | - | $9.979 \times 10^{-06}$ | 0.7452 |

*2) Classifier Performance.*

Performance metrics for the validation set are provided in Table III. Metric values for the distinct set of features, 5 SNPs ($1 \times 10^{-5}$), 32 SNPs ($1 \times 10^{-4}$), 248 SNPs ($1 \times 10^{-5}$) and 2465 SNPs ($1 \times 10^{-2}$) were obtained using optimized F1 threshold values 0.3417, 0.4284, 0.4487 and 0.4594, respectively. Conversely, Table IV shows the performance metrics on the test data when the trained models are used.

*3) Model Selection.*

The ROC curves in Fig. 2 shows the cut-off values for the false and true positive rates using the test set. A clear deterioration in performance as the number of SNPs decrease (P-value threshold increase) can be observed. Note that SNPs with conservative P-value thresholds, are an indication of how significant associations are. In this instance, machine learning demonstrates the limited predictive capacity of highly ranked SNPs when discriminating between case and control samples (obese and non-obese individuals).

## IV. DISCUSSION

GWAS has proven to be useful for identifying common variants with modest to large effects on phenotype. Single SNP scan is still the most extensively used approach. However, in GWAS, SNPs are independently tested for association with some phenotype of interest, ignoring relationships between genetic variants. One important advantage of deep learning is its ability to abstract large, complex and unstructured data into latent representations that can be used to describe SNPs and their interactions. In this study, we have combined bioinformatics tools and techniques with deep learning, as a framework to evaluate the predictive capacity of the most significant SNPs (lowest P-values) extracted from association analysis with logistic regression. Thus, a two-stage methodology has been considered, in which four subsets of loci following quality-control (QC) and association analysis, were selected, and classification tasks were conducted.

QC and logistic regression under an additive genetic model was performed to assess the association between SNPs and binary disease status. Furthermore, instead of selecting the most statistically significant genetic variants based solely on Bonferroni or suggestive levels of associations, we considered SNPs with P-values lower than $1 \times 10^{-2}$, $1 \times 10^{-3}$, $1 \times 10^{-4}$ and $1 \times 10^{-5}$ as shown in Table III and Table IV.

While no SNPs reached genome-wide significant levels of associations (P-value < $2.1 \times 10^{-7}$), five SNPs were suggestive of association (P-value < $10^{-5}$) as shown in Table II. We observed a suggestive association to obesity and the CDH13 gene, with the most significantly suggestive marker being rs763727 (P-value = $1.821 \times 10^{-6}$). The second strongest suggestive association was achieved by the marker rs726553 (P-value = $7.330 \times 10^{-6}$), situated in an intergenic region of the genome. The third suggestive SNP listed in Table II is rs10817737 (P-value = $8.319 \times 10^{-6}$), situated in the TMOD1 gene. Conversely, the suggestive variants rs3050 (P-value = $9.061 \times 10^{-6}$) located in the gene PLEKHG1 and rs1278895 (P-value = $9.979 \times 10^{-6}$) located in an intergenic region. In Fig. 1 these five SNPs are highlighted and labelled. To report the context of the SNPs identified, the Variant Annotator Integrator (VAI) was utilised [32]. This tool reports variants that are within 5000 bases of each transcript. Two of the reported genes have been previously associated with obesity related traits. Particularly, gene CDH13 which has been previously reported to be associated with Adiponectin levels [33], Hypertension [34] and Coronary Artery Disease (CAD) [35]. Additionally, PLEKHG1 has been reported to be associated with blood pressure in African-ancestry populations [36].

TABLE III. VALIDATION SET PERFORMANCE

| P-value | Sens | Spec | Gini | LogLoss | AUC | MSE |
|---|---|---|---|---|---|---|
| $1 \times 10^{-5}$ | 0.9042 | 0.2334 | 0.2537 | 0.6596 | 0.6268 | 0.2338 |
| $1 \times 10^{-4}$ | 0.7394 | 0.6343 | 0.5091 | 0.5867 | 0.7545 | 0.2002 |
| $1 \times 10^{-3}$ | 0.8564 | 0.8502 | 0.8460 | 0.3550 | 0.9230 | 0.1110 |
| $1 \times 10^{-2}$ | 0.9734 | 0.9780 | 0.9862 | 0.1062 | 0.9931 | 0.0271 |

TABLE IV. TEST SET PERFORMANCE

| P-value | Sens | Spec | Gini | LogLoss | AUC | MSE |
|---|---|---|---|---|---|---|
| $1 \times 10^{-5}$ | 0.9717 | 0.0861 | 0.1960 | 0.6791 | 0.5980 | 0.2428 |
| $1 \times 10^{-4}$ | 0.8135 | 0.5024 | 0.4489 | 0.6079 | 0.7244 | 0.2104 |
| $1 \times 10^{-3}$ | 0.9548 | 0.7129 | 0.8242 | 0.3762 | 0.9121 | 0.1189 |
| $1 \times 10^{-2}$ | 0.9604 | 0.9712 | 0.9817 | 0.1150 | 0.9908 | 0.0300 |

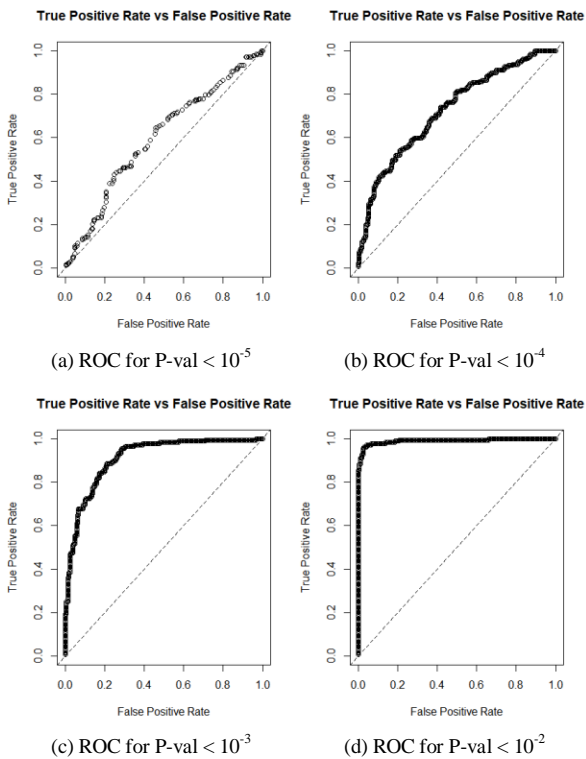

Fig. 2. From (a) to (d) performance ROC curves for test set using trained models with different number of SNPs, according to thresholds: P-value < $10^{-5}$, P-value < $10^{-4}$, P-value < $10^{-3}$ and P-value < $10^{-2}$ respectively.

Although GWAS are useful for identifying common variants of small effect and very rare variants of much larger effect, classification analysis demonstrates that GWAS results fail to classify phenotypes using suggestive or Bonferroni levels of association. Using a deep learning classifier model and genetic variants with P-value < $1\times10^{-2}$ (2465 SNPs) it was possible to obtain the best results (SE=0.9604, SP=0.9712, Gini=0.9817, LogLoss=0.1150, AUC=0.9908 and MSE=0.0300). In contrast, using the 5 suggestive SNPs (P-value < $1\times10^{-5}$) resulted in a significant drop of the results (SE=0.9717, SP=0.0861, Gini=0.1960, LogLoss=0.6791, AUC=0.5980 and MSE=0.2428). The lowest sensitivity (SE=0.0861) value achieved was also reported in the model with 5 SNPs (see Table IV), indicating that model was unable to classify normal individuals correctly. In fact, the models were able to classify risk classes better than normal classes both in validation and test datasets, except when 2465 SNPs ($1\times10^{-2}$) were used. Therefore, the performance of the classifier improved progressively with the number of SNPs included as features. In Fig. 2, as the p-value is increased, an evident deterioration in performance can be observed. This demonstrates that single SNP analysis fails to capture the cumulative effect of less significant variants and their overall contribution to the outcome.

## V. CONCLUSION

The application of GWAS in case control setups have resulted in a plethora of significant genetic variants associated with complex disease phenotypes. However, the predictive capacity of these genetic markers is weak since this approach is based on single-locus analysis, omitting the existence of interactions between loci.

In this study, we have presented a framework for the classification of obesity as a binary phenotype, utilising cases and controls from Geisinger MyCode project. Deep learning classification modelling is applied using four sets of SNPs as features, extracted after association analysis with logistic regression. We achieved the highest results using 2465 SNPs (P-value < $1\times10^{-2}$) as depicted in Fig. 2 (d), where AUC value in test set is 0.9931. Although we have presented encouraging results, the study needs further research to find better ways of identifying reduced sets of genetic variants forming epistasis networks. Recently, stacked autoencoders have been successfully used in preterm delivery data to generate compressed SNP epistatic information which improved classification results [31]. However, mapping SNPs inputs to hidden layers nodes is still a recognized limitation as deep learning approaches act as black box where models become difficult to interpret.

In future work we will extend this study by incorporating stacked autoencoders [37] and association rule mining (ARM) [38]. Combining ARM and the strength of machine learning will help creating a robust method for interpreting deep learning networks.

ACKNOWLEDGMENT

Samples and data in this study were provided by the Geisinger MyCode® Project. Funding for the MyCode® Project was provided by a grant from Commonwealth of Pennsylvania and the Clinic Research Fund of Geisinger Clinic. Funding support for the genotyping of the MyCode® cohort was provided by a Geisinger Clinic operating funds and an award from the Clinic Research Fund. The datasets used for the analyses described in this manuscript were obtained from dbGaP at http://www.ncbi.nlm.nih.gov/gap through dbGaP accession number phs000381.v1.p1.